\input phyzzx.tex

\input epsf

\tolerance=1000

\twelvepoint
\normalbaselineskip=16pt


\REF\Witten{E. Witten, Nucl. Phys. {\it Solutions of Four-Dimensional
Field Theories via M-Theory}, {\bf B500} (1997) 3, hep-th/9703166.}

\REF\HLWb{P.S. Howe, N.D. Lambert and P.C. West, {\it Classical M-Fivebrane
Dynamics and Quantum $N=2$ Yang-Mills}, Phys. Lett. {\bf B418} (1998) 85,
hep-th/9710034.}

\REF\LWa{N.D. Lambert and P.C. West, {\it Gauge Fields and M-Fivebrane
Dynamics}, Nucl. Phys. {\bf B524} (1998) 141, hep-th/9712040.}

\REF\HStelleW{P. Howe, K. Stelle and P. West, {\it A Class of Finite 
Four-Dimensional Supersymmertric Field Theories},
Phys. Lett. {\bf 124B} (1983) 55.}

\REF\Seiberg{N. Seiberg, {\it Supersymmetry and Nonperturbative 
Beta Functions}, Phys. Lett. {\bf B206} (1988) 75.}

\REF\DMNP{P. Di Vecchia, R. Musto, F. Nicodemi and R. Pettorino,
{\it The Anomaly Term in the N=2 Supersymmetric Gauge Theory}, 
Nucl. Phys. {\bf B252} (1985) 635.}

\REF\SW{N. Seiberg and E. Witten, {\it Electomagnetic Duality,
Monopole Condensation and Confinement
in N=2 Supersymmetric Yang-Mills Theory}, Nucl. Phys. {\bf B426} (1994) 19,
hep-th/9407087.}

\REF\DLP{J. Dai, R.G. Leigh and J. Polchinski, 
{\it New Connections Between String Theories}, Mod. Phys. Lett. {\bf A4} 21 
(1989) 2073.} 

\REF\Polchinski{J. Polchinski, {\it D-branes and Ramond-Ramond Charges},
Phys. Rev. Lett. {\bf 75} (1995) 4724, hep-th/9510017.}

\REF\BHOO{J. de Boer, K. Hori, H. Ooguri and Y. Oz,
{\it Kahler Potential and Higher Derivative Terms from M-Theory Fivebrane}
Nucl.Phys. {\bf B518} (1998) 173}

\REF\HY{M. Henningson and P. Yi, {\it
Four-Dimensional BPS-specta via M-Theory}, Phys.Rev. {\bf D57} (1998) 1291,
hep-th/9707251}

\REF\Mikhailov{A. Mikhailov, {\it BPS States and Minimal Surfaces},
Nucl. Phys. {\bf B533} (1998) 243, hep-th/9708068.}

\REF\GLWb{J.P. Gauntlett, N.D. Lambert and P.C. West,
{\it Supersymmetric Fivebrane Solitons}, Adv. Theor. Math. Phys.
{\bf 3} (1998) 1, hep-th/9811024.}

\REF\LWc{N.D. Lambert and P.C. West, {\it Monopole Dynamics from
the M-fivebrane}, hep-th/9811025.}

\REF\AADD{I. Antoniadis, 
N. Arkani-Hamed, S. Dimopoulos and G. Dvali, {\it New Dimensions at 
a Millimeter to a Fermi ad Superstrings at a TeV}, Phys. Lett. {\bf B436}
(1998) 257, hep-ph/9804398.}

\REF\KT{
Z. Kakushadze and S.-H.H. Tye, {\it Brane World}, hep-th/9809147.}

\REF\RSa{
L. Randall and R. Sundrum, {\it Out of this World Supersymmetry Breaking},
hep-th/9810155.}

\REF\BDN{T. Banks, M. Dine and A.E. Nelson, {\it Constraints on Theories with 
Large Extra Dimensions}, hep-th/9903019.}

\REF\Bachas{
C. Bachas, {\it `Desert' in Energy or Transverse Space}, hep-th/9907023.}

\REF\ADS{
I. Antoniadis, E. Dudas and A. Sagnotti, {\it Brane Supersymmetry Breaking},
hep-th/9908023.}

\REF\IQ{
L.E. Ibanez and F. Quevedo, {\it Anomalous U(1)'s and Proton Stability in
Brane Models}, hep-ph/9908305.}

\REF\BBS{K. Becker, M. Becker and A. Strominger,
{\it Fivebranes, Membranes and Non-Perturbative String Theory},
Nucl. Phys. {\bf B456} (1995) 130, hep-th/9507158.}

\REF\HLWa{P.S. Howe, N.D. Lambert and P.C. West, {\it The Three-brane
Soliton of the M-Fivebrane}, Phys. Lett. {\bf B419} (1998) 79,
hep-th/9710033.}

\REF\Bateman{H. Bateman, {\it Higher Transcendental Functions}, vol. 2, 
McGraw-Hill, New York, 1953.}

\REF\Gog{M. Gogberashvili, {\it Gravitational Trapping for Extra
Dimensions}, hep-ph/9808347.}

\REF\RS{L. Randall and R. Sundrum, {\it An Alternative to 
Compactification}, hep-th/9906064.} 

\REF\ADDK{N. Arkani-Hamed, S. Dimopoulos, G. Dvali and
N. Kaloper {\it Infintely Large New Dimensions}, hep-th/9907209.}

\REF\BS{ 
A. Brandhuber and K. Sfetsos {\it Non-Standard 
Compactifications with Mass gaps and Newtons Law}, hep-th/9908116.}

\REF\HLW{P.S. Howe, N.D. Lambert and P.C. West, {\it The Self-Dual
String Soliton}, Nucl. Phys.  {\bf B515} (1998) 203,
hep-th/9709014.}

\REF\BBMOOY{K. Becker, M. Becker, D.R. Morrison, H. Ooguri, Y. Oz and Z. Yin
{\it Supersymmetric Cycles in Exceptional Holonomy Manifolds and Calabi-Yau 4-F
olds}, Nucl. Phys. {\bf B480} (1996) 225, hep-th/9608116.}

\REF\GP{G.W. Gibbons and G. Papadopoulos, 
{\it Calibrations and Intersecting Branes}, hep-th/9803163.}

\REF\AFS{B.S. Acharya, J.M. Figueroa-O'Farrill and B. Spence, 
{\it Branes at Angles and Calibrated Geometry}, hep-th/9803260.}

\REF\Barbon{J.L.F. Barbon, {\it Rotated Branes and N=1 Duality},
Phys. Lett. {\bf B402} (1997) 59, hep-th/9703051.}

\REF\HOO{K. Hori, H. Ooguri and Y. Oz,
{\it  Strong Coupling Dynamics of Four-Dimensional N=1 Gauge Theories
from M Theory Fivebrane}, Adv. Theor. Math. Phys. {\bf 1} (1998) 1,
hep-th/9706082.}

\REF\Wittenb{E. Witten, {\it Branes and the Dynamics of QCD}, Nucl.
Phys. {\bf 507} (1998) 658, hep-th/9706109.}

\REF\LWb{N.D. Lambert and P.C. West, {\it D-Branes in the Green-Schwarz
Formalism}, hep-th/9905031.}

\REF\HSW{P. S. Howe, E. Sezgin and P. C. West, {\it Covariant Field
Equations of the M-Theory Fivebrane}, Phys. Lett. {\bf B399} (1997) 49,
hep-th/9702008.}

\REF\HS{P. S. Howe and E. Sezgin, {\it D=11, p=5},
Phys. Lett. {\bf B394} (1997) 62,
hep-th/9611008.}

\REF\Berger{
M.Berger, 
{\it Sur les Groupes D'holomony des Varieties a Connexion Affine et des 
Varieties Riemanniennes}, Bull. Soc. Math. France {\bf 83} (1955) 279; 
{\it Remarques sur le Group D'holonomie des Varieties Riemannienes},
C.R.Sci. Paris {\bf 262} (1966) 1316,  
for a review see S. Salamon, {\it Riemannian Geometry and Holonomy Groups}, 
Longman 1989.}

\REF\GLWa{J.P. Gauntlett, N.D. Lambert and P.C. West,
{\it Branes and Calibrated Geometries}, Comm. Math. {\bf B202} (1999) 571, 
hep-th/9803216.}

\REF\VA{D.V. Volkov and V.P. Akulov, 
{\it Is the Neutrino a  Goldstone particle}, 
Phys. Lett. {\bf B46} (1973) 109.}

\pubnum={KCL-TH-99-36\cr hep-th/9908129}
\date{August 1999}

\titlepage

\title{\bf M-Theory and Hypercharge}

\centerline{N.D. Lambert}

\centerline{and}

\centerline{P.C. West\foot{lambert, pwest@mth.kcl.ac.uk}}
\address{Department of Mathematics\break
         King's College, London\break
         England\break
         WC2R 2LS\break
         }

\abstract
We discuss the possibility that the electro-weak and strong interactions
arise as
the low energy effective description of branes in M-theory.
As a step towards constructing such a model we
show how one can naturally obtain $SU(N_1)\times SU(N_2)\times U(1)$
gauge theories from branes, including matter in the bi-fundamental
representation of $SU(N_1)\times SU(N_2)$ which are fractionally
charged under $U(1)$.



\endpage


\chapter{Introduction}

Branes occur as solutions to supergravity theories in ten and eleven
dimensions and play a natural role in the dynamics of string theories
and the underlying,  but elusive,  M-theory.
From this origin it was somewhat
surprising to find that branes had the ability
to describe sophisticated
properties of four-dimensional  quantum field
theories that had been studied for many years.
These quantum field theories quantum field
theories  live on the intersection of branes and
their dynamics, at low energies,  can be computed from the dynamics of
the branes.   The prototypical example is the spontaneously broken
$SU(2)$ N=2 Yang-Mills theory in four dimensions which can be realised
on the intersection of NS-fivebranes and D-fourbranes in type IIA string
theory. At strong coupling this configuration is lifted to M-theory
and appears as a single M-fivebrane wrapped on a Riemann surface [\Witten].
Moreover the complete low energy
effective action can be computed from the  classical dynamics of the
M-fivebrane [\HLWb,\LWa] and this agrees precisely with the perturbative 
[\HStelleW,\Seiberg,\DMNP] and non-perturbative [\SW] quantum dynamics
of the N=2 Yang-Mills theory.

Given a self intersecting or wrapped
M-fivebrane we can compute from the
its equations of motion an effective action which describes the
low energy
behaviour of the configuration.  This effective action
lives on the self intersection of the M-fivebrane and its  degrees of
freedom are the moduli corresponding to the M-fivebrane soliton solution that
describes the wrapping. The procedure has some
similarities to that used to compute the motion of monopoles except
now the fields of the underlying theory describe the embedding of the
M-fivebrane and the behaviour of its worldvolume gauge field. Since
M-theory is the strong coupling description of the type IIA string
we can, by
shrinking one of the dimensions in the background spacetime 
in which the M-fivebrane
is wrapped, find a corresponding  IIA description. 
In this limit the single M-fivebrane becomes a system of
intersecting NS-fivebranes and D-fourbranes. The D-fourbranes have associated
with them a non-Abelian gauge theory, the degrees of freedom of which
arise as the modes of open strings stretched between the D-fourbranes
[\DLP,\Polchinski]. 
Since  these modes also describe the
fluctuations of the branes their  behaviour provides a description of
the the brane configuration in the IIA limit which is the strong
coupling limit of that found from the M-fivebrane equations of motion.

An infinite number of the modes of the open strings
have masses of the order of the string scale, but  there exist a finite
number of modes that  are either massless or have masses on the scales
set by the brane configuration. At energies below the string scale only
these  latter modes are relevant and one can consider the effective
action that results from the scattering of the open strings and
describes their behaviour. The terms in this action that involve
parameters of positive mass dimension will resemble a conventional
renormalisable quantum field theory which we refer to as the dual theory.

For the case of $N=2$
 Yang-Mills theory mentioned  above the complete low energy effective
action of the dual quantum field theory including all
perturbative and non-perturbative effects agrees with the low
energy effective  action found from the M-fivebrane equations of
motion. However, one would not expect  that the higher
derivative corrections to agree and indeed it is
known that for
$N=2$ gauge theories the higher derivative terms of Yang-Mills theory
and those from the M-fivebrane dynamics differ [\BHOO,\LWa].
While one may not expect the low energy effects of the dual theory
and the low energy effective action of the M-fivebrane to agree
in all circumstances one would expect a substantial  degree of agreement at
energies below the scales of the brane configuration. In particular these two
theories share the same spectrum of states and gauge symmetry. However
if a microscopic description of M-theory were available then we would expect
that the brane effective field theory and the dual Yang-Mills effective
field theory would agree at scales below the Plank scale.

The M-fivebrane is particularly interesting to study since it has a two-form
gauge field in its worldvolume theory which satisfies a non-linear
self-duality constraint. There is no known microscopic
description available  as there is for  D-branes, however  the low energy
effective dynamics of M-fivebranes are known, although  only for the case
of a single M-fivebrane corresponding to an Abelian two-form. Now since
M-theory is the strong coupling limit of type IIA string theory, the
M-fivebrane describes the strong coupling limit of D-fourbranes, including
quantum effects. Thus the two-form gauge field contains information
about strongly coupled non-Abelian D-fourbrane dynamics, even though
the only low energy effective actions that
one obtain at present from the M-fivebrane  contain
only Abelian gauge fields.
However these Abelian theories do
incorporate some  of the non-Abelian aspects of the
dual theory, up to the characteristic symmetry breaking
scale. For example, the non-linearities in the effective action
reflect the scattering of the  low energy modes which result from the
interactions of the dual theory determined by non-Abelian gauge
symmetry, in a manor similar to how the four-Fermi theory  accounted for the
interactions of the Standard Model.  It is also possible to consider
M-fivebrane configurations
corresponding to confined gauge theories, with no long range gauge group.
Although the lack of a microscopic description makes quantitative
predictions difficult to obtain. In fact, the degrees of freedom
that occur in the dual theory that  have masses of the order of scales
set by the brane configuration also have a description in terms
of the dynamics of the M-fivebrane. For example  the
the solitonic solutions that occur in the underlying non-Abelian
theory, in particular monopoles. In Yang-Mills theories these rely  on
the non-Abelian structure  for their existence  and do not occur as
solutions of the low energy effective action. However their properties
can be inferred from the brane dynamics by introducing M-twobranes
which intersect the M-fivebrane worldvolume 
 [\Witten,\HY,\Mikhailov,\GLWb,\LWc].  In this way
the brane configurations mimic the behaviour of the dual Yang-Mills
theories.

If String Theory is to play a role
in the complete description
of physics then it  must contain the physics of the
Standard Model at low energy.
It seems very
natural that branes will play a part in the way the Standard Model
arises (for some recent work and some reviews see 
[\AADD,\KT,\RSa,\BDN,\Bachas,\ADS,\IQ] 
and the references therein).
In these scenarios it is usually assumed
that the Standard Model itself is
the low energy effective action of String Theory, perhaps
with some background brane configuration.
Moreover it is usually assumed that
the minimal supersymmetric extension of the Standard Model, or a grand unified
extension of it, is the
low energy action of String Theory, which should  become valid at some
scale near $M_W$. In this paper
we will take a significantly
different approach to deriving
low energy effective dynamics of the electro-weak
interactions. We will exploit the fact that a single M-fivebrane
 naturally leads to a low energy effective action  with broken
supersymmetry, spontaneously broken gauge symmetry whose only
unbroken gauge groups are $U(1)$ factor or are confinement.

There is of course very  good experimental
confirmation that the electro-weak and strong nuclear force
is described by the $SU(3)\times SU(2)\times U(1)$ quantum field
theory of the Standard Model, at least  up to energy scales
around its symmetry breaking scale $M_W$. As we have
explained above, although branes naturally incorporate spontaneously
broken gauge symmetry and supersymmetry breaking they do not
easily lead to effective theories with unconfined non-Abelian gauge
groups. On the other hand, in Nature the only unbroken gauge groups are the
confined $SU(3)$ and the  $U(1)$ of electromagnetism.
Since branes
naturally lead to effective theories with these groups, in this paper
we explore the possibility that M-fivebrane dynamics can produce
a low energy  effective theory with the above unbroken gauge groups
and that this low energy effective theory will have a dual
theory that is the Standard Model, e.g. that this theory has the same
spectrum of states as the Standard Model. As we have discussed above this
brane derived low energy effective theory will not agree exactly with all
the dynamical predictions of the Standard Model, but one may hope that they
will be sufficiently close to agree with the parts of the Standard
Model which have been verified experimentally. If so such a brane
configuration would hold out the exciting  possibility of new,
braney physics beyond the $M_W$ scale. We also hope that the
work we describe here will be useful for studying other
phenomenological models derived from branes.

Although in this paper we will not find such a brane derived low
energy effective action we will systematically identify the problems
that one encounters and illustrate how some of them may
be solved within the context of a toy model. These problems can be
thought about in essentially two ways. Firstly we must find a
brane configuration with the correct modes (e.g. the dual theory
should be the Standard Model).
Secondly we must analyse this configuration
in M-theory to obtain the brane low energy effective action
(e.g. we need to determine the geometry of the M-fivebrane and
its resulting zero modes).

One of the most obvious problems is to find a brane configuration which
has the gauge group $SU(3)\times SU(2)\times U(1)$. A particular difficulty
arises because states in the Standard Model are charged under
all three  gauge groups, whereas in the D-brane description
of branes, states can only be charged under two groups -
one for each end point of an open string. A way out of this dilemma is to
recall that the brane gauge
group that arises from open string theory is $U(N)$,
rather than $SU(N)$.
In previous studies of gauge theories from branes the additional
$U(1)$ factors of $U(N)$ are either trivial and decouple,
or they are frozen out of the low energy dynamics.
In this paper we  will provide explicit examples which rely on
compactified spaces and show that this is generically not the case.
Furthermore we derive a specific formula for the
$U(1)$ charge of state in terms of its representation under
$SU(N_1)\times SU(N_2)$. Intriguingly  this formula
reproduces the correct hypercharges of many fields in the Standard
Model, although it also contains fields with the opposite
hypercharges.

In section two we discuss how one can find dynamical $U(1)$ gauge groups
in the low energy brane theory. We also
present a toy brane configuration with
four-dimensional $N=2$ supersymmetry whose dual  theory
is an $SU(3)\times SU(2)\times U(1)$ gauge theory. This model has
``quarks'' states that transform in the $({\bf 3},{\bf 2})$ of
$SU(3)\times SU(2)$ and have $U(1)$ charge $\pm1/6$.
In section three we present a systematic discussion of the problems and
issues in constructing a more realistic brane configuration.
Section four provides a brief conclusion of our work.

\chapter{Hypercharge from Branes}

In this section we wish to construct type IIA brane configurations whose
low energy description is a 
four-dimensional $N=2$ gauge theory where
some residual $U(1)$ gauge fields do not decouple. 
Our strategy is to consider intersecting M-fivebranes in eleven dimensions
where $x^{10}$ is compact with period $2\pi R$.
In particular we consider configurations which have four-dimensional
Poincare symmetry  and
preserve $N=2$ supersymmetry, i.e. threebrane solitons on the M-fivebrane. 
Furthermore we wish to consider M-fivebrane
intersections which, when $R\rightarrow 0$ and dimensionally reduced to 
type IIA string theory in ten dimensions, 
can be interpreted as configurations of 
NS-fivebranes with D-fourbranes
suspended between them, as first studied in [\Witten]. 
This then allows us to approach
their analysis in two ways. On the one hand we can use perturbative
type IIA D-brane physics to calculate the low energy modes and determine
the corresponding dual Yang-Mills theory. On the other hand we may  
solve these models in M-theory where they appear as a single
M-fivebrane wrapped on a surface $\Sigma$ and the classical
equations of motion provide a good approximation.

It was noted in [\Witten] that one can avoid freezing out all the 
$U(1)$ factors if the brane configuration is compactified in the $x^6$ 
direction, although only configurations for which the residual $U(1)$
was trivial were considered. In this section we wish to extend this analysis
to obtain non-trivial $U(1)$ gauge groups in the dual theory and 
evaluate their effective action.
Therefore we will generalise the so-called Elliptic models in [\Witten],
where $x^6\cong x^6 + L$.  These are configurations of
M-fivebranes which intersect over the four
dimensions $x^\mu$, $\mu=0,1,2,3$. Alternatively
they can be interpreted as a single M-fivebrane wrapped on a 
surface $\Sigma$ embedded in ${\bf R}^2\times{\bf T}^2$.  Demanding that
half of the sixteen supersymmetries are preserved requires that $\Sigma$
is a Riemann surface [\BBS,\HLWa].

We introduce the complex coordinate $s = x^6+ix^{10}$ and the Weierstrass 
$\wp$-function $\wp(s)$ associated to the torus
$$
s \cong s + \omega_1\ ,\quad s\cong s+\omega_2\ ,
\eqn\torus
$$
with $\omega_1 =L$ and $\omega_2 = 2\pi iR$. As is well-known $\wp$ is a 
two-to-one map of
the flat torus onto ${\bf CP}^1$ 
(in other words it is a one-to-one map onto a the twice cut ${\bf CP}^1$) 
and satisfies the cubic equation 
$$
(\wp(s)')^2 = 4 (\wp(s)-e_1)(\wp(s)-e_2)(\wp(s)-e_3)\ ,
\eqn\wpdef
$$
where  $e_i=\wp({\omega_i\over2})$ 
for $i=1,2,3$ and $\omega_3 = \omega_1+\omega_2$. 
In this case the limit $R\rightarrow 0$ corresponds to 
going to type IIA string theory on ${\bf R}^{9}\times S^1$ and the 
limit $L\rightarrow\infty$ corresponds to uncompactifying this 
$S^1$ factor.
Our next step is to introduce a variable $t$ through
$$
t = \wp(s)\ .
\eqn\tdef
$$
The role of $\wp(s)$ generalises that of $e^{-s/R}$, which maps the 
cylinder to ${\bf CP}^1$ in the non-elliptic models 
of [\Witten].

To analysis our configurations we will consider the limit 
$L\rightarrow\infty$. In this limit $\wp(s)$ takes the form (see for example 
[\Bateman])
$$
\wp(s) = {1\over R^2}\left[{1\over 12} 
+ {1\over e^{-s/R}+e^{s/R}-2}\right]\ .
\eqn\wplimit
$$
Note that two of the branch points $e_1 = \wp({\omega_1\over2})$ and 
$e_3=\wp({\omega_3\over2})$ have coalesced 
to ${1\over 12R^2}$, whereas the other two
branch points are $e_2 = \wp({\omega_2\over 2}) = -{1\over 6R^2}$
and $e_4 = \infty$. 
It will be helpful to perform a modular transformation on $t$ so as 
to map the branch points
$e_2$ and $e_4$ to $-1$ and $+1$ respectively and the degenerating
branch points $e_1$ and $e_3$ to $\infty$. Thus we introduce
$$
\tilde t = {t +{5\over 12R^2}\over t - {1\over 12R^2}}\ ,
\eqn\ttdef
$$
so that, in the limit $L\rightarrow\infty$, 
$\tilde e_4=-\tilde e_2=1$ and $\tilde e_1=\tilde e_3=\infty$.

To describe an embedding
of the M-fivebrane we need to specify a function
$$
F(s,z)=0\ ,
\eqn\Fdef
$$
where $z= x^4+ix^5$ is a complex coordinate on the M-fivebrane and
$s$ is a complex coordinate transverse to the M-fivebrane.
As shown in [\BBS,\HLWa], 
this will preserve four-dimensional $N=2$ supersymmetry
for any holomorphic function $F$.
A natural form for the embedding of the surface
$\Sigma$ into ${\bf R}^2\times {\bf T}^2$ is
$$
A_{1}(z)\tilde t^k 
+ A_{2}(z)\tilde t^{k-1}
+\ldots +A_{k+1}(z)=0\ ,
\eqn\embedding
$$
where $A_i(z)$ are polynomials of degree $N_i$ in $z$.

We may now justify this form for the embedding by  considering  
the limit $L\rightarrow\infty$ where \embedding\ reduces to
$$
A_{1}(z)(e^{-s/R}+e^{s/R})^k
+A_{2}(z)(e^{-s/R}+e^{s/R})^{k-1}+\ldots + A_{k+1}
=0\ ,
\eqn\curve
$$
which is the  form of the curve  used in [\Witten] for non-elliptic models. 
Note that $s$ only appears in the combination $s^{-s/R}+e^{s/R}$ 
so that these brane configurations
all have the symmetry $s\leftrightarrow -s$. This is not unexpected from 
our construction  since, when viewed as
limit of a configuration which is periodic in $s$, the behaviour as
$s\rightarrow\infty$ and $s\rightarrow-\infty$ must be the same.
Clearly, for finite $L$, the 
semi-infinite D-fourbranes going off to the left are 
identified with those going off to the right.

Let us consider the perturbative type IIA description of this configuration
obtained in the $R\rightarrow 0$ limit. After compactification 
components of the
M-fivebrane that are wrapped on $x^{10}$ become D-fourbranes 
in ten dimensions,
whereas the components which are not wrapped on $x^{10}$ 
become NS-fivebranes. Since
$\wp(s)$ is a two-to-one map the surface described
by \embedding\ consists of a $2k$-sheeted cover of the $z$ plane
with some number of branch cuts on each sheet which is determined by
the degrees of the polynomials $A_i(z)$. In the type IIA
limit each sheet corresponds to an NS-fivebrane while each branch cut
appears as a D-fourbrane stretching between two NS-fivebranes.  
It is helpful to now consider
the limit $L\rightarrow\infty$. Due to  $A_1(z)$, there
will be $N_1$  D-fourbranes attached to the first NS-fivebrane and $N_1$
D-fourbranes attached to the last NS-fivebrane which stretch to
$s = -\infty$ and $s=\infty$ respectively. When $L$ is finite these
semi-infinite D-fourbranes are identified with each other and become $N_1$ 
finite D-fourbranes stretched between the
first and last NS-fivebrane. Note that, and this will be crucial in our 
analysis, we do not restrict 
attention to conformal models with the same number of D-fourbranes
suspended between all pairs of NS-fivebranes as was studied in [\Witten].
 
The low energy description of this system is obtained by analysing open
strings with end points on the D-fourbranes. The NS-fivebranes are 
infinitely heavy in the weak coupling limit so that their dynamics are
suppressed. Strings that begin and end on the same set of parallel 
D-fourbranes give a five-dimensional vector multiplet with sixteen 
supersymmetries and gauge group
$U(N_i)$. However, the presence of the NS-fivebranes causes the $x^6$
direction of the D-fourbrane to be compactified and projects out half of
the supersymmetries and states. This leaves us with a four-dimensional
$N=2$ $U(N_1)\times U(N_2)\times \ldots \times U(N_k)$ 
gauge theory. We must
also consider open strings with one end point on the $i$th set of D-fourbranes
and one end point on the $j$th set of D-fourbranes with $i\ne j$. 
These strings give an $N=2$
hyper multiplet in four dimensions in the bi-fundamental representation of
$U(N_i)\times U(N_j)$ [\Witten].

For example consider the simplest case $k=1$ where
$$
t =\wp(s)={1\over 12R^2}\left({A_2/A_1-5\over A_2/A_1+1}\right)\ .
\eqn\simplest
$$
In the $L\rightarrow\infty$ limit this
is simply
$$
A_1(z)(e^{-s/R})^2 + 2A_2(z) (e^{-s/R}) + A_1(z) = 0 \ ,
\eqn\example
$$
which corresponds to two NS-fivebranes with $N_2$ D-fourbranes suspended
between them with $N_1$ semi-infinite D-fourbranes coming off to the left
and $N_1$ semi-infinite D-fourbranes coming off to the right of the 
NS-fivebranes. For finite $L$ the semi-infinite D-fourbranes are identified
and become finite so that we have a 
four-dimensional $N=2$ $U(N_1)\times U(N_2)$ gauge theory with a single 
hypermultiplet in the bi-fundamental representation.

We now  need to discuss the low energy dynamics of these configurations 
obtained from M-theory. 
In static gauge the bosonic fields of the M-fivebrane 
worldvolume theory consist of 
five scalars which correspond to its position in the 
transverse space and a two-form gauge field whose three-form field
strength satisfies a non-linear self-duality constraint.
In our case all but two of the scalars are trivial and can be set to zero.
The remaining two scalars can be identified with the complex coordinate $s$.
To obtain the effective action for the brane configuration we need only
expand the equations of motion to second order in field strengths and
spacetime derivatives $\partial_\mu$.
>From the constraints of $N=2$ supersymmetry it is sufficient to consider
only the scalar zero modes since the rest of the low energy effective action
may be determined uniquely from the purely scalar terms. To this end we
follow the procedure used in [\HLWb], however for more realistic cases
with less supersymmetry one must use the analysis of the
three-form given in [\LWa]. When the
three-form is set to zero the lowest order term in the effective Lagrangian 
is [\HLWb] 
$$\eqalign{
{\cal L} &= {1\over2}\int d^2 z \partial_\mu s\partial^\mu \bar s\ ,\cr
&= {1\over8}\int d^2 z {\partial_\mu t\partial^\mu\bar t \over
|(t-e_1)(t-e_2)(t-e_3)|}
\ .\cr}
\eqn\effaction
$$
To obtain the low energy dynamics of the soliton defined by \embedding, 
we write the $A_i(z)$ polynomials in \embedding\ as
$$
A_i(z) = u_{i,0} z^{N_i} + u_{i,1} z^{N_i-1} +\ldots +u_{i,{N_i}}
\eqn\polydef
$$ 
and let the moduli $u_{i,a}$ become functions of the four 
coordinates $x^\mu$, $\mu=0,1,2,3$. 
The Lagrangian for these scalars can be evaluated as
$$
{\cal L} = {1\over 8}\int d^2 z \sum_{i=1}^{k+1}\sum_{a=0}^{N_i}
{\partial_\mu u_{i,a}\partial^\mu\bar u_{i,a} \over
|(t-e_1)(t-e_2)(t-e_3)|}\left| {\partial t\over \partial u_{i,a}}
\right|^2\ .
\eqn\Lagrangian
$$

Let us now concentrate on the simplest configuration \simplest\ 
corresponding to a $U(N_1)\times U(N_2)$ gauge theory and 
write
$$\eqalign{
A_{1} &= u_0 z^{N_1}+u_1z^{N_1-1}+\ldots+u_{N_1}\ ,\cr
A_{2} &= v_0 z^{N_2}+v_1z^{N_2-1}+\ldots+v_{N_2}\ .\cr}
\eqn\polys
$$
Here we will be interested in cases where $N_1 <N_2$  and we may
therefore  rescale $z$ and the curve
\simplest\ so that  $u_0=v_0=1$. Recall that the zeros of $A_i(z)$
represent the positions of the $N_i$ parallel D-fourbranes [\Witten]. 
In particular the sub-leading coefficients $u_1$ and $v_1$ are the
sum of the positions of two sets of parallel D-fourbranes.

Let us now check the convergence of the integral over $z$. For $L<\infty$ the
branch points $e_1,e_2$ and $e_3$ are distinct and there are  two
potential divergences: from the large $z$ limit and from points where
$\partial t/\partial u_{i,a}$ diverges. For large $z$ we see that 
$t \sim 1/12R^2$ whereas 
$\partial t/\partial u_{i,a} \sim z^{-(N_2-N_1+1)}$
so that the integrand is well behaved. 
Next we see that 
$\partial t/\partial u_{i,a}$  diverges when
$A_2=-A_1$ or when $\partial ({A_2\over A_1})/\partial u_i^a$ diverges.
In the first case $t$ also diverges and this causes the integrand
in \Lagrangian\ to vanish at these points. The second case can  occur when
$z\rightarrow\infty$ which we have already considered and potentially at 
$A_1=0$. However when $A_1=0$ one finds that  
$\partial t/\partial u_{i,a}$ is finite.
Thus we find that
all the  moduli $u_1,...,u_{N_1}$ and $v_1,...,v_{N_2}$ 
in $A_1(z)$ and $A_2(z)$ have finite action
for finite $L$. 
Thus, in contrast to the non-elliptic case, the centre of mass of the 
D-fourbranes has finite energy. Therefore the $u_1$ and $v_1$ moduli
will appear in the low energy effective action as non-trivial 
dynamical fields.

Clearly for any configuration we may
shift $z \rightarrow z+z_0$ without changing the Lagrangian $\cal L$.
Under this shift the centre of mass moduli are altered according to
$$
u_1\rightarrow u_1+N_1z_0\ ,\quad v_1\rightarrow v_1 + N_2z_0\ . 
\eqn\weight
$$
It follows that the low energy effective action has a 
non-trivial dependence on the relative centre of mass between 
the two sets of D-fourbranes
$$
u_{R}={u_1 \over N_1} - {v_1 \over N_2}\ ,
\eqn\urdef
$$
but depends trivially on the over-all centre of mass
$$
u_{C}={u_1+v_1\over N_1+N_2} \ .
\eqn\ucdef
$$
Therefore although the 
low energy effective Lagrangian will contain kinetic terms for both
$u_C$ and $u_R$,  $u_C$ will be a free field.  
In addition, due to $N=2$ supersymmetry,
the low energy effective action must also have two $U(1)$ gauge fields
$$\eqalign{
A^R_{\mu}&={1\over N_1}A^1_\mu - {1\over N_2}A^2_\mu\ ,\cr
A^C_{\mu}&={1\over N_1+N_2}A^1_\mu+{1\over N_1+N_2}A^2_\mu \ ,}
\eqn\Adefs
$$
which are the $N=2$ superpartners of $u_R$ and $u_C$ at linearised level
respectively [\LWa]. However since $u_C$ is a free field $A^C_{\mu}$ 
completely decouples from the dynamics and appears trivially in the effective
action.
On the other hand the effective action
will depend non-trivially on $u_R$, corresponding to massive states which do
carry $A^R_\mu$ charge that have been integrated out. Therefore our
$U(N_1)\times U(N_2)$ brane configuration has a non-trivial  
$SU(N_1)\times SU(N_2)\times U_R(1)$ gauge theory on its worldvolume
and we may ignore the fields $u_C$ and $A^C_\mu$.

Let us now examine the charges of the various states under the 
$U_1(1)\times U_2(1)$ factor of 
$U(N_1)\times U(N_2)\cong U_1(1)\times U_2(1) \times SU(N_1)\times SU(N_2)$. 
The charge of states from an open string
stretched between two distinct D-branes is $(1,-1)$. Here the relative 
minus sign comes from the orientation
of the open string. CPT conjugate states in the $N=2$ multiplet 
are obtained by considering an open string with the opposite orientation. 
If an open string begins and ends on the same
D-brane then it must be   
uncharged since it is indistinguishable from an open string with the 
opposite orientation, i.e. it carries $U_1(1)\times U_2(1)$ charge
$(0,0)$.
Thus for an open string with one end on a set of $N_1$ parallel D-fourbranes
and the other on a set of $N_2$ parallel D-fourbranes the 
$U_C(1)$ charge always vanishes; $Q_C=0$. However the $U_R(1)$ charge is
$$
Q_R = {1\over N_1} - {1\over N_2}\ .
\eqn\hypercharge
$$
From this formula we see that  if an open string begins and ends on
the same set of D-fourbranes then its $U_R(1)$ charge is zero, as one
expects since it is in the adjoint representation of $U(N)$. If we consider
the conformal cases where $N_1=N_2$ then the strings stretched 
between different sets of D-fourbranes are also uncharged under
$U_R(1)$. In these cases the $U_R(1)$ is therefore trivial and can be 
omitted from the low energy dynamics as in the models of [\Witten]. 
However in general open
strings from one set of D-fourbranes to a different set will yield states
with fractional $U_R(1)$ charge. 

From the point of view of the M-fivebrane each of the low energy $U(1)$
vector gauge fields comes from the period of the Abelian two-form gauge
field over a one-cycle in $\Sigma$ [\LWa]. However
the self-duality constraint of the three-form field strength implies that
a one-cycle and its conjugate one-cycle of the Riemann surface give
rise to the same low energy $U(1)$. So if $\Sigma$ has genus $g$, we
expect $g$ $U(1)$ gauge fields in the low energy effective action.
It is easy to see the extra $U(1)$ we have just discussed is associated
with the one-cycle corresponding to $x^6\cong x^6 + L$. 
Thus for a more general embedding with $k>1$ we expect only one additional
$U(1)$, corresponding to an 
$SU(N_1)\times SU(N_2)\times\ldots \times SU(N_k)\times U(1)$ gauge theory, 
since by compactifying $x^6$ we increase the genus of $\Sigma$,
and therefore the number of $U(1)$ gauge fields, by one. 
This was also noted in [\Witten] from a different perspective. To be more
explicit we have shown that the one-cycle $x^6\rightarrow x^6+L$ 
of the background spacetime coincides with a one-cycle of the embedded
surface $\Sigma$. This leads to an additional $U(1)$ gauge field in the
low energy effective theory which in the type IIA limit may be identified
with the relative centre of mass of the D-fourbranes.

We should comment here on an apparent difficulty: if  
a brane low energy effective action has $n$ $U(1)$ gauge fields 
how do we know whether a particular one of these $U(1)$'s
arises from the Cartan subalgebra of $SU(N)$ of whether it
arises as a relative $U_R(1)$? To be more concrete suppose we have a 
model with two $U(1)$ gauge fields in the low energy effective
action. How can we tell if this comes from spontaneously broken $SU(3)$
gauge theory or an $SU(2)\times U(1)$ theory? As one can see from above
analysis, the appearance of the $U_R(1)$ factor 
is associated with the non-trivial
topology of spacetime, i.e. compactness of $x^6$. This means that there
is no point in the brane moduli space where this non-trivial one-cycle
associated to $U_R(1)$ degenerates and a non-Abelian symmetry is restored.
In contrast for an $SU(3)$ gauge theory there will be points in the brane
moduli space, i.e. vacuum expectation values of the scalars, where the
gauge symmetry will be enhanced and the corresponding one-cycles in $\Sigma$
are degenerate.

Thus we can construct a toy model for the electro-weak interactions 
by taking $k=1$, $N_1=2$ and $N_2=3$, i.e. we consider the embedding
$$
(z^2+u_1z+u_2)\tilde t + (z^3+v_1z^2+v_2z+v_3)=0\ .
\eqn\toy
$$
The corresponding type IIA brane configuration is shown in figure
one. The vertical lines denote the NS-fivebranes, the horizontal lines
the D-fourbranes and the dashed lines the various possible open strings
that can stretch between the D-fourbranes.
The non-trivial
part of the low energy dynamics consists of an $N=2$ supersymmetric
$SU(2)\times SU(3)\times U_R(1)$ 
gauge theory. In addition we have a hyper multiplet of ``quarks''
in the $({\bf 2},{\bf 3})$ of $SU(2)\times SU(3)$ with $U_R(1)$ charge
$\pm 1/6$. In fact, since the open strings that stretch between the two sets
of parallel D-fourbranes can wrap $n$ times around the compact $x^6$
direction, we find an infinite tower of quark ``generations'' whose
masses increase linearly with $n$. Of course this model is far from
describing the quarks of the Standard Model. Most notably it is non-chiral,
has $N=2$ supersymmetry and the $SU(3)$ factor of the
gauge group is spontaneously broken, rather than confined. Nevertheless 
this model arises naturally and quite simply from brane configurations and
the charges of the ``quarks'' are realistic.


\midinsert
\epsfysize= 5cm
\centerline{\epsffile{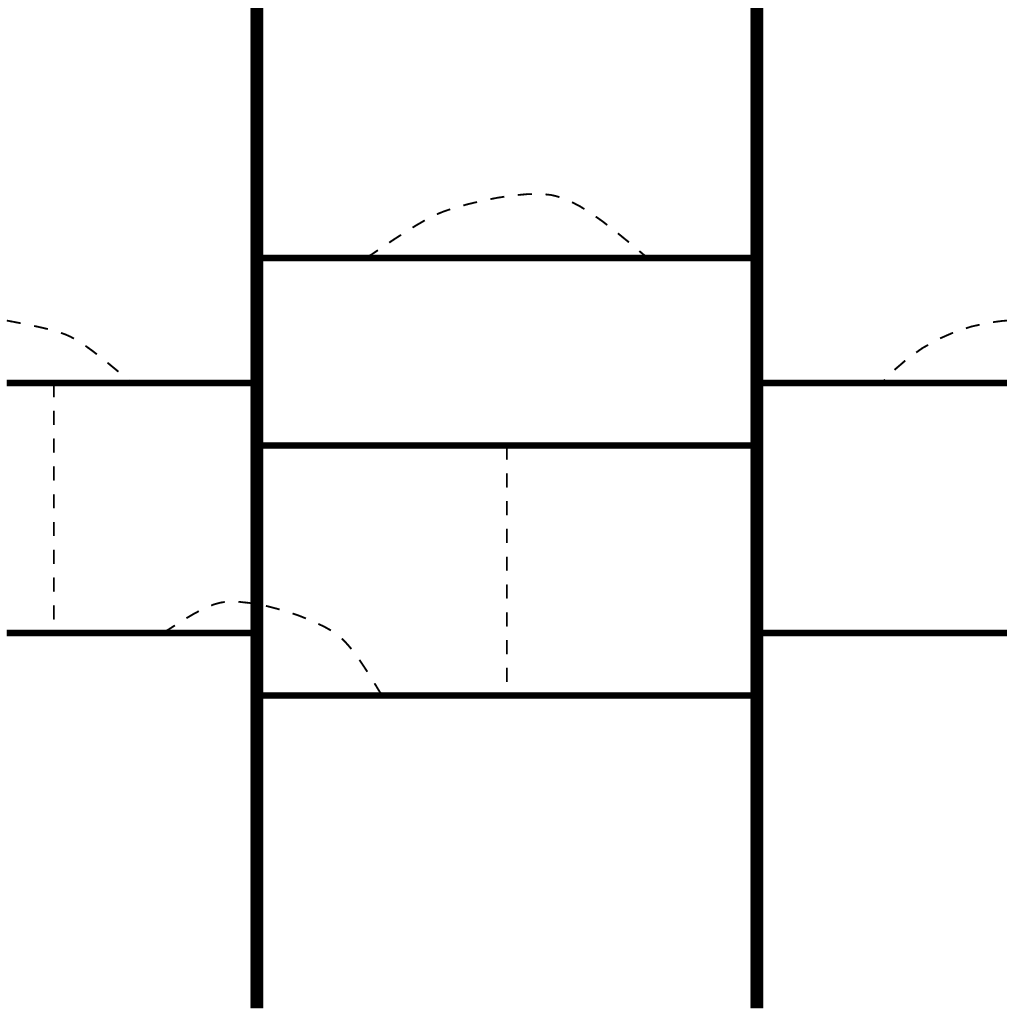}}
\centerline{\it Figure 1}
\endinsert

\chapter{Towards a Fivebrane Realisation of the Electro-Weak and Strong
Interactions}

In this section we wish to explore the possibility that M-theory
can  describe
the electro-weak and strong interactions. More precisely
we wish the theory of   electro-weak and strong interactions to
arise from an intersecting M-fivebrane configuration in an
eleven-dimensional background spacetime.
 The effective low energy action can be
calculated from M-theory and will live on the worldvolume of the
self-intersection of the M-fivebrane. This space is to be identified
with the world in which we live and so is  taken to be four
dimensional Minkowski space ${\bf M}^{1,3}$.  The remaining
two dimensions of the M-fivebrane are wrapped on a two dimensional
manifold $\Sigma$ embedded in  the background spacetime. This effective
action  will describe the  electro-weak and strong interactions. A
necessary condition for this to occur is that the dual gauge theory
that arises in this brane configurations as identified in the IIA limit
must be a
$SU(3)\times SU(2)\times U(1)$ gauge  theory which has the same matter
content as the Standard Model. In particular the
resulting brane configuration that
emerges in the IIA limit must have the field gauge group
$SU(3)\times SU(2)\times U(1)$ which is spontaneously broken to
$U(1)$ and a confining $SU(3)$ and have the
field content of the Standard Model. Of course 
it must therefore have no supersymmetry.

The background spacetime  must
be a solution of  eleven dimensional supergravity of the form
${\bf M}^{1,3} \times Q$, where $Q$ is
a seven dimensional space. The M-fivebrane dynamics describe the
embedding of $\Sigma$ into  $Q$. Here we will use the
classical equations of motion of the M-fivebrane to determine features
of  the low energy effective action for this embedding.
These are only a good approximation when the manifolds
$Q$ and $\Sigma$ are sufficiently large and smooth. Although when
considering branes that lead to confining gauge groups this is not
always the case,
the classical equations of motion along with what is known about the
quantum corrections to the M-fivebrane  will
provide a qualitative description of the low energy dynamics. From the
perturbative IIA string theory point of view, for which there is the
dual Yang-Mills  theory interpretation, an analysis of
the classical M-fivebrane dynamics incorporates the
strong coupling regime of this theory. The limit to the IIA
theory is found by shrinking one of the one-cycles in the background
spacetime. Clearly in order to obtain a system of NS-fivebranes and
D-fourbranes this cycle must belong to both $\Sigma$ and $Q$.
Thus our task is to identify a suitable choice for
 $\Sigma$ and $Q$ that will lead to a low energy
effective action for the M-fivebrane
which describes the electro-weak and strong interactions.

In order to illustrate more clearly the problems that must be solved
let us first consider the toy model discussed in the previous
section. In this case we consider a background  spacetime
$Q= {\bf T}^2\times {\bf R}^2\times {\bf R}^3$ where the ${\bf R}^3$
factor is trivial. The M-fivebrane is wrapped around a Riemann surface
 $\Sigma$ embedded in
${\bf T}^2\times {\bf R}^2$ defined by the embedding \toy.
The self-intersection of
the M-fivebrane in spacetime appears,
from the perspective  of the M-fivebrane, as the presence of threebrane
soliton solutions on its worldvolume.
Using the  classical M-fivebrane dynamics we can compute
the low energy effective action for the motion of these threebranes.
Indeed the scalar part of this low energy effective action is given by
\effaction.
The result is a low energy effective action in four-dimensional
Minkowski space. More precisely, the degrees of freedom of this low
energy effective action arise from moduli in  the solution
of  the M-fivebrane equations of motion.  In the
threebrane solution only the transverse scalar coordinates of  the
M-fivebrane are active and changes of the moduli that only affect these
coordinates correspond to deformations of the
embedded  surface on which the M-fivebrane is wrapped.  These
moduli lead to scalar degrees of freedom in the low energy
effective action, namely $u_R,u_2,v_2,v_3$.
There are also moduli which correspond to a
non-vanishing worldvolume two-form gauge field  and
are associated  with large gauge
transformations. Such moduli arise as periods of the two-form potential over
one-cycles on $\Sigma$ and lead to vector fields in
the four-dimensional low energy effective action [\LWa].
For example in the toy model there are four non-trivial one-cycles:
two from the set of three parallel D-fourbranes, one from the set of
two parallel D-fourbrane and the fourth comes from the $x^6$ cycle in
spacetime. This leads to an $U(1)^4$ vector gauge potential in the low energy
effective dynamics. There  are four conjugate one-cycles on the
Riemann surface, however due to the self-duality constraint on the
three-form field strength these do not lead to additional low energy
vector fields (this will be discussed in greater detail below). 
Finally we have
Fermionic moduli that  lead to spin 1/2 particles in the low energy
effective action.

As we explained above, we can consider the  limit in
which one dimension $x^{10}$ of the background torus is
shrunk to zero. In this limit  the wrapped
M-fivebrane becomes  a system  of intersecting NS-fivebranes
and D-fourbranes in type IIA string theory.
The excitations of this brane configuration are described by the
open  strings  stretched between the D-fourbranes. The modes of the
open strings can be divided into those that have masses on the
string scale, and are therefore rather heavy, as well as a finite
number of modes whose  masses are much below the string scale.
The latter include massless modes for  open strings whose ends are
attached to the same D-fourbrane or whose ends are fixed to D-fourbranes
that touch at one or more points, and massive  modes for those open
strings whose ends  are fixed to D-fourbranes that are never  touch. The
mass of the latter is proportional to the minimum length of the
stretched open string and so are fixed by the scales of the brane
configuration, rather than the string scale.
The lowest energy modes of these open strings
correspond to the excitations  of the brane configurations. At a first
approximations the dynamics of these modes, which are in fact given by
open string perturbation theory, can be
described  by the dual Yang-Mills quantum field
theory on the worldvolume of the
intersection. Thus using the  connection between
M-theory and IIA string theory we can  identify the dual
quantum field theory.
In addition the complete low energy effective  action is exactly
described by the low energy dynamics for the self-intersecting
M-fivebranes [\HLWb,\LWa].

Although  the
non-linearity  of the low energy effective action can be thought of
as arising  from integrating out  particles such as the non-Abelian
gauge fields it is difficult to infer the existence and properties of
these particle from the low energy effective action.
However,
all the particles of the quantum field theory on the brane, as
identified in the IIA limit,  do occur naturally in the M-theory
description. 
The  particles that occur in the IIA limit
that have masses on brane scales   arise
as  excitations of the 
string stretched between different D-fourbranes. In M-theory, since there are
only M-twobranes and M-fivebranes, when we lift a IIA
configuration the open strings between
the D-fourbranes  must correspond to  M-twobranes which end on
the M-fivebrane. These M-twobranes
appear on the  worldvolume of the M-fivebrane as a self-dual string 
solitons that carry the two-form gauge field charge [\HLW]. 
For general configurations,  these states  occur solitons on the M-fivebrane
with  non-vanishing worldvolume gauge field corresponding to 
a self-dual string  wrapping $\Sigma$ [\GLWb,\LWc].
The type of
particle one finds depends on how the self-dual string wraps itself
around $\Sigma$ [\Witten,\HY,\Mikhailov]. 
In this way one can find in particular
the  the charged spin one vector bosons $W^{\pm}$ and the monopoles of
the $SU(2)\times SU(3)$ factor. Such M-fivebrane solutions were explicitly
constructed in reference [\GLWb,\LWc] 
with in the context of the M-fivebrane
wrapped on a Riemann surface embedded in flat spacetime in the case
of a single $SU(2)$ gauge group. Although
the M-fivebrane equations only described a $U(1)$ gauge theory
it was argued in [\LWc] that finite energy solutions
corresponding  to the
non-Abelian vectors and monopoles do exist.
The low energy scattering of these
particles can also be deduced from the M-fivebrane equations of
motion by letting the moduli associated with the solution (including
the self-dual string) become dynamical [\LWc].

Thus despite the rather different techniques used to find the
two descriptions, one from M-theory and the other from the
perturbative  IIA string,  there is
a one-to-one correspondence between the spectrum and low energy
dynamics of the two theories. Although we have been describing systems
with $N=2$, where there are significant constraints, the qualitative
features of the dual quantum Yang-Mills theory and M-fivebrane gauge theory
should agree for systems with less and even no supersymmetry.

Let us now
return to the task in hand, to construct a solution of M-theory that
describes the electro-weak and strong interactions. We require a
worldvolume threebrane soliton, described by
the wrapping of the M-fivebrane on a surface
$\Sigma$ embedded in a background spacetime of the form
$M^{1,3}\times Q$.
Our first restriction on $Q$ arises because  gravity is not a low energy
mode of the brane and in fact propagates in the full eleven-dimensional
spacetime. The simplest way to obtain a $1/r^2$ force law for gravity,
rather than a $1/r^9$ law, is  compactify all directions
$x^4,x^5,x^6,\ldots,x^{10}$ so that gravity is four-dimensional
on macroscopic scales. Therefore we  assume that $Q$ is a
compact manifold.  Note that some alternatives where the extra
dimensions are non-compact 
have recently appeared [\Gog,\RS,\ADDK,\BS].

Since we wish to find a description of the electro-weak and strong
nuclear forces the   M-fivebrane wrapped on $\Sigma \subset Q$ should
break all supersymmetries. There are two ways in which this could
happen. Either the background spacetime
$Q$ breaks all the supersymmetries of M-theory,
or the background preserves some or all of the supersymmetries of
M-theory, but the two-cycle
$\Sigma$ is not supersymmetric (i.e. is not a calibrated submanifold 
[\BBS,\BBMOOY,\GP,\GLWa,\AFS]).
In either
case one must ensure that the configuration is stable.
In the later case  the preservation of some supersymmetry by
the background space-time implies that the manifold   will
possess reduced holonomy. If the background gauge field
is zero such manifolds are classified [\Berger]
and the only seven-dimensional manifolds which do not
have a direct product structure have $G_2$ holonomy.  The
residual supersymmetry is then  broken by the wrapping of the
M-fivebrane  on
$\Sigma$.

A systematic study of the possible supersymmetric 
intersections of a single M-fivebrane was given in reference [\GLWa]. 
If the compact manifold $Q$ has a
direct product structure then there are a number of ways of wrapping
the M-fivebrane such that supersymmetry is preserved. One example
being given in the previous section. However, if $Q$ does not have a
direct product structure then  there are no background manifolds
with two submanifolds
$\Sigma$ such that any supersymmetry is preserved by the wrapped
M-fivebrane. In particular, $G_2$ holonomy manifolds generically 
possess nonsupersymmetric two-cycles. Even though wrapping a M-fivebrane on 
such a  two-cycle dos not preserve supersymmetry there
may exist in the background manifold  cycles of different dimensions
which wrapping the M-twobranes
or other M-fivebranes leads to supersymmetric states. For example,
$G_2$  holonomy manifolds do possess
non-trivial three and four-cycles over which one can wrap 
M-fivebranes without breaking all the supersymmetry [\GLWa]. 
These would lead to
supersymmetric states in spacetime  corresponding to topological
defects in ${\bf M}^{1,3}$. 
We observe that if we require $Q$ to not possess a direct
product structure and the self intersection of  the M-fivebrane to be 
four-dimensional then the
resulting low energy effective action will break supersymmetry.

Once one has identified a suitable type IIA brane configuration and
its corresponding lift to an embedding $\Sigma \subset Q$ the next step
to analyse the low energy effective dynamics and states using the
M-fivebrane dynamics. We now describe these two steps in greater detail.

\subsection{The IIA Limit}

As is well known the Standard Model is based on a non-supersymmetric
$SU(3)\times SU(2)\times U_Y(1)$ Yang-Mills
gauge theory which is spontaneously broken to $SU(3)\times U(1)$
and in addition the $SU(3)$ factor is confined. The matter content
consists of three generations, which,
from a group theoretic point of view are identical but whose masses
vary dramatically. Here we shall concentrate on the lightest generation
which has the following the matter content
$$
\matrix{ && SU(2)&  SU(3)  & Y\cr
&& & &\cr
q_L &&  {\bf 2} &{\bf 3} & 1/6\cr
e_L && {\bf 2} &{\bf 1} &  -1/2\cr
u_R && {\bf 1} &{\bf 3} &  2/3\cr
d_R && {\bf 1} & {\bf 3} &  -1/3\cr
e_R &&  {\bf 1} &{\bf 1} & -1\cr
}
$$
where $L/R$ denotes the left/right-handed Weyl component of the
spinor  field.

As we have discussed above, it is
rather straightforward to find brane configurations that lead to the
gauge groups $SU(N_1)\times SU(N_2)$; one simply takes $N_1$ and $N_2$
parallel D branes which are suspended between NS-fivebranes. The additional
$U(1)$ factor which one might naively expect to be present is
 frozen out if one demands that
the low energy effective action  is  finite [\Witten]. One way to avoid such an
infinity and so find a
$U(1)$ factor is to compactify directions in which the D-fourbranes lie
in. This occured in the so called elliptic models of [\Witten].
Unfortunately  in these  models considered there the gauge group was
$SU(N)^k\times U(1)$ and the $U(1)$ gauge
field decoupled from the rest of the theory.

In the previous section we
showed that one could compactify one of the directions and find a
theory with the  gauge
group $SU(3)\times SU(2)\times U(1)$ and the hypermultiplets are
charged under the $U(1)$ factor. We recall that the existence of an
additional $U(1)$ relies on the presence of a one-cycle of $\Sigma$
that coincides with a one-cycle in the background space $Q$. 
Although this model possess $N=2$
supersymmetry and so is unrealistic it clearly illustrates that one
can find the correct gauge group. Let us consider its matter content
and postpone
the problem that these states form a non-chiral $N=2$ hyper multiplet. The
toy model
possessed  a   ``quark''  that was in the same
group  representation  as $q_L$, including the correct $U(1)$ charge (it 
also possessed a ``quark''
with the opposite $U(1)$ charge). In addition the toy
model had a generation structure, although  these states had a linearly
increasing mass as a function of their wrapping number $n$, whereas in
the Standard Model the mass of the successive generations increases
much more  rapidly.

The toy model can be made more realistic in a number of ways.
To gain a confining $SU(3)$ we can
follow [\Barbon,\HOO,\Wittenb]
and rotate one of the NS-fivebranes so that it lies in
the $x^7,x^8$ plane rather than in the  $x^4,x^5$ plane. From the
geometrical point of view this corresponds
to embedding $\Sigma$ non-trivially
in six of the dimensions of $Q$.
In these configurations
there the distance between the NS-fivebranes which connect the three
parallel D-fourbranes are no longer constant and therefore the three parallel
D-fourbranes will lie along the shortest distance between them. In this
case the M-fivebrane worldvolume is no longer a smooth manifold but has
a singularity where the D-fourbrane sit on top of each other.
The M-fivebrane equations of motion are no longer valid and in addition
the system can no longer be described by a single M-fivebrane.
The D-fourbranes no longer have any scalar moduli and lead to a
confining $SU(3)$ [\HOO,\Wittenb].
Indeed if $Q$ is a curved  manifold then one might expect there to be
no scalar moduli for a particular embedding since in general
the minimal sized cycles and distances will be unique, i.e. there will be no
flat directions in the M-fivebrane moduli space. To remove the massless
scalar from the broken $SU(2)$ sector of the theory one may suppose that
$Q$ is curved so that there are two paths which are (locally) the shortest
distance between the NS-fivebranes. This would then remove the massless 
scalar mode that represents the fluctuations of the D-fourbranes but still lead
to a spontaneously broken $SU(2)$.

We can also gain a more realistic matter content  by introducing
additional sets of D-fourbranes into the toy model. In particular we can
add a  single D-fourbrane stretched between additional NS-fivebranes
to our configuration in a manner similar to figure two. Although the
configuration in figure two is not of the form described in
the previous section, it might still be possible to find other embeddings
which allow it to be compactified. From the
discussion in the last section this would not lead to additional $SU(N)$
or $U(1)$ factors of the gauge group. However open strings stretched
from the additional D-fourbrane
to the sets of two and three parallel D-fourbranes we would obtain
states in the $({\bf 1},{\bf 2})$ and $({\bf 3},{\bf 1})$
of $SU(3)\times SU(2)$. Their representation would be the same as for
$e_L$ and $u_R$ respectively,
again  obeying the $U(1)$ charge formula \hypercharge\
derived in the previous section (up to a sign).
On the other hand it is not clear how to obtain states corresponding to
$d_R$ and $e_R$. In this case the  formula \hypercharge\ does not
appear to work for these states, although it only differs by an
additive factor of one.


\midinsert
\epsfysize= 5cm
\centerline{\epsffile{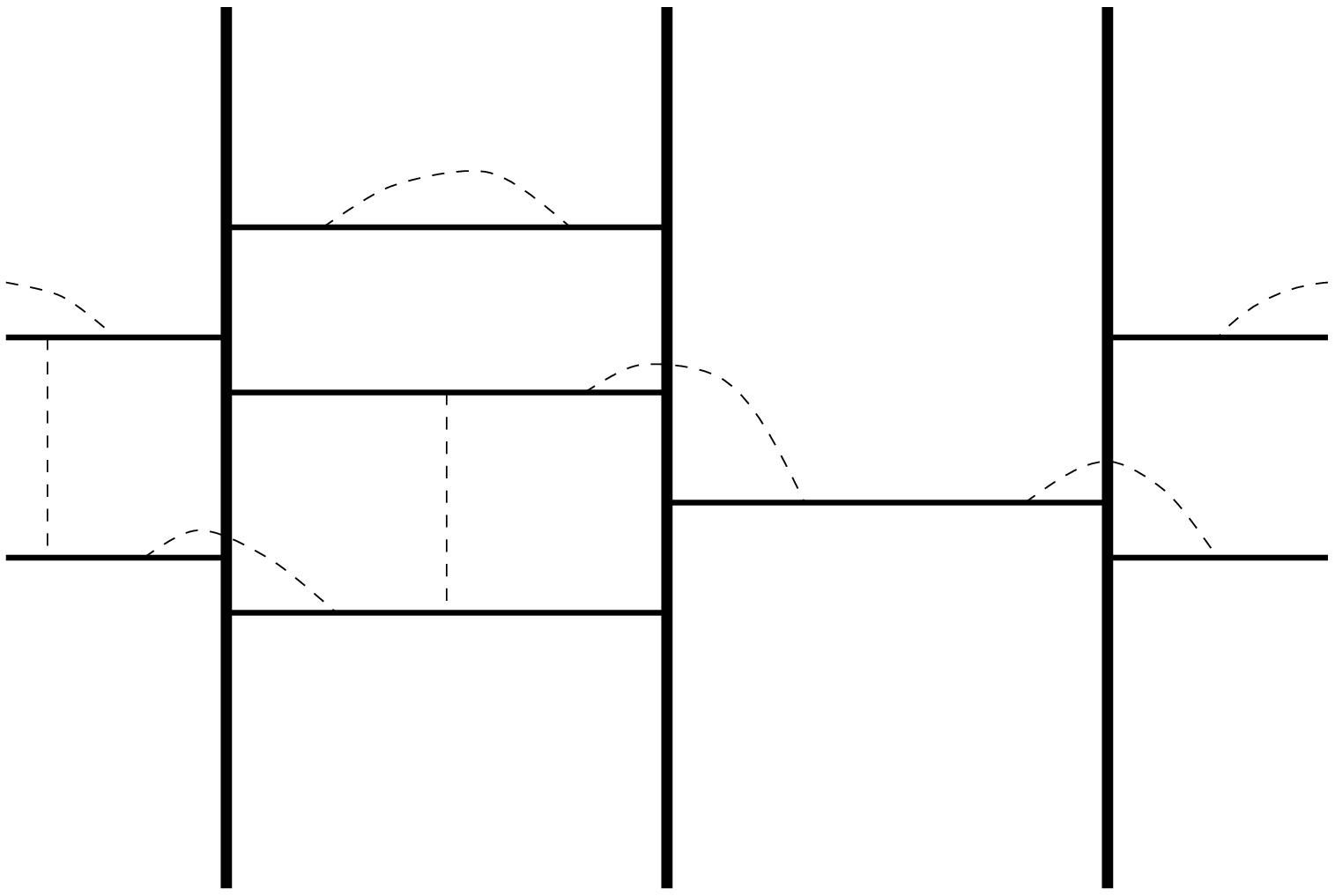}}
\centerline{\it Figure 2}
\endinsert

Another key problem with the toy model is that it is non-chiral. The open
strings that are used to study a D-brane configuration naturally come
with sixteen supersymmetries and produce four-dimensional $N=4$
supermultiplets. By placing NS-fivebranes into the configuration
the number of preserved supersymmetries is decreased and the size of
the supermultiplet is reduced accordingly [\Witten]. In particular, within the
Green-Schwarz formalism, this leads to a mechanism where as
supersymmetries are projected out and the multiplet is reduced [\LWb]. In
some cases,  when  all of
the supersymmetries are projected out, only the highest spin component
of the open string
multiplet survives [\LWb]. Thus we might expect that only the
spin 1 and spin 1/2 states of the vector and hyper multiplets
respectively survive in many
non-supersymmetric brane configuration. This is just as is required to make
contact with the fields of the Standard Model, i.e. vector fields with
chiral matter.

\subsection{The low energy effective action }

Let us now consider the requirements that the M-fivebrane wrapping on
the two-cycle must possess in order to describe the required low
energy effective theory.

One of the remarkable features of
the Standard Model  is that there are no explicit mass terms  in the
theory as  a consequence of the different multiplet structures of the
Weyl left-handed and  right-handed Fermions, the gauge symmetry for the
vector bosons and the doublet structure of the Higgs. After the
symmetry breaking all particles get a mass through the Higgs mechanism
except for the photon, gluons  and neutrinos.
Particles which get their mass in the symmetry breaking
must be regarded as associated with the symmetry breaking scale and
will not appear as dynamical  moduli in the low energy theory.
Hence we should only find moduli in the M-fivebrane embedding solution
that  correspond to the photon, gluons and the neutrino. However as
we mentioned above the gluons, due to confinement, are not described by
the classical low energy effective action of the M-fivebrane. Instead one
needs to know the microscopic theory underlying the M-fivebrane dynamics
describe them.  All the other
particles  in the Standard Model should
arise as more complicated M-fivebrane
solutions consisting of a self-dual string wrapping around the
two-cycle $\Sigma$. As we have discussed above these
particles arise as modes of  the stretched open strings between the
D-fourbranes in the type IIA limit.

It is instructive to examine the M-fivebrane equations that must be
solved. We use the covariant equations  of motion found in
[\HSW] which were derived from the
superembedding formalism applied to the M-fivebrane [\HS].
Neglecting the background three-form gauge field
the scalars of the obey the equation [\HSW]
$$
G^{mn} \nabla_m (\partial_n X^{\underline p})=0\ ,
\eqn\scalareqn
$$
where the covariant derivative $\nabla _m$ is a connection with
respect to worldvolume and background indices. Here and below underlined
indices run over the eleven-dimensional background space $\underline m,
\underline n,\underline p=0,1,2,...,10$, whereas
$m,n,p,...=0,1,2,...,5$ are worldvolume coordinates.
To be precise
$$
\nabla _m V_n ^{\underline p}= \partial _m V_n ^{\underline p}
-\Gamma_{mn}^q V_q ^{\underline p}
+\partial _m X^r \Gamma _{\underline r \underline s}^{\underline p}
 V_n^{\underline s}\ ,
\eqn\connection
$$
where the connections are the standard Christoffel
connections coefficients
with respect to the induced metric $g_{mn}$ on the worldvolume and
background metric $\gamma_{\underline m\underline n}$ respectively (note that
the last term was neglected in [\HSW]).
We also need to introduce tangent indices for the M-fivebrane worldvolume
$a,b,c,...$ and the vielbein $g_{mn} = e_m^{\ a}e_n^{\ b}\eta_{ab}$.
The tensor $G^{mn}$ is defined as
$$\eqalign{
G^{mn}&=(m^2)^{ab}e_a^{\ m}e_b^{\ n}\ ,\cr
m_a^{\ b} &= \delta_a^{\ b} - 2h_{acd}h^{bcd}\ .\cr}
\eqn\Gdef
$$
Here $h_{abc}$ is related to the three-form field strength and will be
defined below.

The actual embedding is given by a field configuration in which the
worldvolume gauge field $B_{mn}$ vanishes,
the flat coordinates $X^\mu$ of the self-intersection of the
M-fivebrane are taken to belong to four-dimensional
Minkowski space;
$\partial _\mu X^{\underline p}=0, \ \mu=0,1,2,3$. Equation \scalareqn\
then determines the behaviour of the scalar coordinates on the
two remaining coordinates $x^4,x^5$
of the M-fivebrane and specify how it  wraps around the
two-cycle $\Sigma$ of the background space $Q$.

The scalar moduli of the embedding solution become dynamical
when they are allowed to depend on the flat worldvolume
which has coordinates $x^\mu$ and their equation of motion can be
derived from equation \scalareqn\ by dimensionally reducing the
equations of motion over $x^4$ and $x^5$.
The behaviour of these moduli can be studied
in the absence of worldvolume gauge fields. In this case the
equation minimises the volume of $\Sigma$.
The smallest volume is likely to be given by
the M-fivebrane wrapping tightly to any fixed background cycles.
In such a
case one will find no scalar moduli as any deviation will lead to a
solution with an increased energy. This is a welcome feature as we
should find no scalar fields in our low energy effective action.

It is instructive to  realise the difference with the toy model
studied in section two where scalar moduli were present. In this
case the background space had certain flat directions, partly because
there was no fixed two-cycle in $Q$ around which the M-fivebrane
wrapped.
The equations of motion were satisfied by any Riemann
surface $\Sigma$ and
thus there were corresponding scalar moduli parameterising
the moduli space of
Riemann surfaces with a fixed genus. If $Q$
had a more complicated topology
and geometry then one would not
expect to find moduli spaces of solutions
with the same energy, but rather a fixed surface $\Sigma$.

Let us now examine how gauge fields can arise in the effective
action.
The worldvolume three-form  field strength $H=dB$ of the M-fivebrane
obeys a non-linear self-duality constraint. This is most easily
described by first considering
a linearly self-dual field $h$
$$
h_{abc}= {1\over 3!}\epsilon_{abcdef}h^{def}\ .
\eqn\hdef
$$
Next we construct
the field strength $H_{abc}$ as
$$
H_{abc} = (m^{-1})_a^{\ d}h_{dbc}\ .
\eqn\Hhdef
$$
A final step is to impose that $H$ is closed, so that we may identify
$H=dB$. As a
consequence of this non-linear self-duality constraint the
closure of $H$ is in fact equivalent to the equation of motion [\HSW]
$$
G^{mn} \nabla_m H_{npq}=0\ .
\eqn\Heqn
$$

In general it is rather complicated to solve this
system of equations, but to discover the presence  of vector fields in
the low energy effective action it is sufficient to work to
linearised order in $H$ and  to zeroth order in derivatives of the
scalar with respect to the coordinates $x^\mu$. In this case
we need only solve $H=\star H$ and  $dH=0$.
We may write $H$ in the form
$$
H= {\cal F} \wedge \omega +k\wedge \Omega + L
\eqn\Hlinearised
$$
where
$\omega$ and $\Omega$ are one-forms and two-forms on $\Sigma$
respectively and $k$, ${\cal F}$ and $L$ are one, two  and three-forms 
on ${\bf M}^{1,3}$ respectively.  
To the approximation to which we are working, the
equation
$dH=0$ implies that
$$
d_2 \omega=0\ ,\quad d_4 {\cal F}=0\ ,\quad  d_4 k=0\ ,\quad d_4L=0\ ,
\eqn\linearisedeqn
$$
where $d_2$ and $d_4$ are the exterior derivatives on $\Sigma$ and
${\bf M}^{1,3}$ respectively.
Low energy modes arising from  $k$ and $L$ contribute additional  
four-dimensional scalars and thus we must also ensure that these 
do not occur.

The self-duality of $H$ then implies that
$$
H_{\mu\nu i}={1\over 2}{\rm det}(e)\epsilon _{\mu\nu\rho \kappa}
\epsilon_{ij}  g^{jk} H_{\rho \kappa k j}\ ,
\eqn\lineraisedsd
$$
where $i,j=4,5$.
We may decompose $\omega$ into  the self-dual one-forms
$\omega_A^{\pm}$ which satisfy $\omega_{Ai}^{\pm}=
\pm i{\rm det} e\epsilon_{ij}  g^{jk} \omega_{Ak}^{\pm}$,
$A=1,2,3,...,g$. Writing
$$
{\cal F}\wedge\omega
= \sum_{A=1}^g\left({\cal F}_A^+ \omega_A^{+}
+ {\cal F}_A^- \omega_A^{-}\right)\ ,
\eqn\Fdecomposition
$$
we find that
the above self duality condition becomes
$$
{\cal F}^{\pm}_{A\mu\nu}= \pm{i\over 2}
\epsilon _{\mu\nu\rho\kappa}{\cal F}_A^{\pm\rho\kappa}\ .
\eqn\Fsd
$$
Defining ${\cal F}_A^{\pm}=F_A\pm i\star F_A$ we conclude that $F_A$
is real
and satisfies $d F_A=0$ and $d\star F_A=0$. As such it represents a
$U(1)$ gauge field in the low energy effective action.
Hence we may conclude that the for every pair of conjugate two-cycles in
$\Sigma$,  we find a $U(1)$ gauge field in the
effective action.  For our purposes  we require only one gauge field
hence only one pair of one-cycles. We recall from section two that
the existence of a $U(1)$ factor required a one-cycle in $\Sigma$
that was also a one-cycle in the background spacetime $Q$.

Let us now turn our attention to the worldvolume Fermions
$\Theta^{\underline \gamma}$, $\underline \gamma = 1,2,3,...,32$.
Without gauge fixing these obey the equation [\HSW]
$$
g^{mn}\partial_ n \Theta ^{\underline
\gamma} (1-\Gamma)_{\underline
\gamma}^{\ \underline
\beta } (\Gamma_{m})_{\underline \alpha \underline \beta }
=0\ .
\eqn\Fermion
$$
where $E_{\underline n}^{\ \underline a}$ is the eleven-dimensional
vielbein,
$\Gamma_{ m}=\partial _m X^{\underline n} E_{
\underline n}^{\ \underline a}\Gamma_{\underline a}$ and
$$
\Gamma = {1\over 6!\sqrt{-g}}\epsilon^{mnpqrs}\Gamma_{mnpqrs}
+{1\over 3} h^{mnp}\Gamma_{mnp}\ .
\eqn\gammadef
$$
Note that
this equation involves a Dirac operator that is
quite different to that usually found for Fermions propagating in a curved
background. As
such the usual discussions of  zero modes which are for example found
in Kaluza-Klein theories will not apply in an obvious way.
The Fermionic moduli are the zero modes of \Fermion\ in the
presence of the threebrane solution. The Fermions that appear in the
low energy effective action are found by letting these zero modes
depend on $X^\mu$ and their equation of motion is derived by
from \Fermion. It is immediately clear that these Fermions couple to
the worldvolume background gauge field strength and do not couple
minimally to the gauge field. Since the worldvolume gauge field
contains all the gauge fields of the effective action, it follows that
the zero mode Fermions of the effective action are uncharged under
the $U(1)$ gauge fields of the effective action.

For our application we require only one left-handed Fermion, that is
the neutrino and two other similar particles for the other
generations. The neutrino is uncharged under the unbroken $U(1)$
and is therefore  compatible with it being a  zero mode  of
\Fermion.
As the toy model discussed above illustrates
the other, charged,  particles of the
Standard Model should
appear in M-theory as solitonic solutions that generalise the
embedding solution to include a non-vanishing worldvolume gauge
field. The appearance of these states as stretched open strings
in the IIA limit ensures the existence of
such solutions on the M-fivebrane.

A final problem that we will mention here concerns the masses of the states.
As we mentioned states corresponding to matter fields arise a soliton
solutions on the M-fivebrane and therefore it is not unreasonable that
one could find a spectrum of masses that have a complicated pattern.
However, one should also explain the large hierarchy of masses found
in the Standard Model which extends over
many  orders of magnitude even below $M_W$. 
In particular, the electron should appear as
just such a soliton, however, one must  explain why its mass
is six orders of magnitude lower than the mass of the W-bosons.
However there is an additional problem in 
the toy model with the gauge fields. Namely
although the toy model has the correct gauge group it has  four low
energy $U(1)$ vector fields whereas the Standard Model has only one.
If the $SU(3)$ factor is made confining, as we discussed above, then
this would still leave two
$U(1)$ vector fields from the Cartan subalgebra of $SU(2)\times U(1)$. 
Therefore we need to find some mechanism whereby
only a linear combination of these two $U(1)$ vector fields remains
massless.

According to  the ideas set out in this
paper the  observable massless fields of the Standard Model, namely
the neutrino and the photon, arise as moduli of the M-fivebrane soliton
soliton associated with the Fermion and worldvolume gauge field
respectively. However, the Fermions of the M-fivebrane are the
Goldstone Fermions corresponding to the breaking of the  supersymmetry
of M-theory by the M-fivebrane. As such the
neutrino is a Goldstone Fermion. This is reminiscent of the old
suggestion of Volkov and Akulov [\VA] where the neutrino is a Goldstone 
Fermion
corresponding to the breaking of supersymmetries 
in four-dimensional spacetime. Although in our case the
broken supersymmetries are not those associated to our 
four-dimensional spacetime. Similarly, here 
one may regard the photon as a
Goldstone boson which is probably related to the breakdown of
the central charge symmetries  that occur in the eleven-dimensional
supersymmetry algebra.

The remaining particles in the Standard Model  all carry electric charges
and must arise as solitons associated with self-dual strings on the
fivebrane which are wrapped on the surface $\Sigma$. The charge of
these particles arises from the charge carried by the self-dual
string. Their mass is
given by the rest energy of the soliton and is unlikely to be
zero. Indeed, since the massless zero modes of the M-fivebrane are necessarily
uncharged with respect to the final low energy $U(1)$ gauge field, 
there is a link between the  electric charge of a
particle and its mass. This generalises similar  mass inequalities 
which occur when
masses arise in spontaneous symmetry breakdown or when BPS bounds are
saturated.

\chapter{Conclusion}

We have discussed a novel way to derive electro-weak and strong
interactions from M-theory and String Theory. In particular we have
shown how one can naturally obtain a non-trivial
$U(1)$ hypercharge in the low energy effective dynamics of branes.
Although we have not
presented a concrete model with the correct features of the
Standard Model,
we have outlined the problems that one faces and also some possible
means to over come them.
In particular we have shown how the
$SU(3)\times SU(2) \times U(1)$ gauge group
arises naturally from branes
and with matter that has realistic hypercharge assignments.
Of course even if a model which satisfies our  criteria is
found, it would be remarkable if it were to be in
agreement with the vast amount of
experimental evidence on electro-weak and strong interactions.
However, if
this is the case then such a model would hold out the very exciting
possibility of significant new physics that could test M-theory and String
Theory
in the next generation of particle accelerator experiments.
We also
hope that the discussion we have given here will be helpful
for other phenomenological applications of branes.
We note that the model we described here appears
to be significantly harder
to use to calculate cross sections and scattering than a standard quantum
field theory. Its main advantage is
that it is a manifestly finite string theoretic, and so unified,
description of Nature that arises in a simple and straightforward manner
from M-theory.

\chapter{Acknowledgements}

One of the authors (PCW) would like to thank the Universities of
Tasmania and Naples for their hospitality where some of this work
was carried out and also the Royal Society for a travel grant to Tasmania 

\refout

\end